\documentclass[preprint,showpacs,preprintnumbers,amsmath,amssymb]{revtex4}
\usepackage{graphicx}
\usepackage{dcolumn}
\usepackage{bm}

\begin{document}


\preprint{APS/123-QED}

\title{Class of exact solutions of the Skyrme and the Faddeev model \\}

\author{Minoru Hirayama}
 \email{hirayama@sci.toyama-u.ac.jp}
\author{Jun Yamashita}
 \email{fxjun@jodo.sci.toyama-u.ac.jp}
\affiliation{Department of physics, Toyama University, Toyama 930-8555, {\bf Japan}\\}


\date{today}

\begin{abstract}
Class of exact solutions of the Skyrme and the Faddeev model are presented. In contrast to previously found solutions, they are produced by the interplay of the two terms in the Lagrangians of the models. They are not solitonic but of wave character. With an appropriate choice of field variables, the field equations of the two models are written in exactly the same form. The solutions supply us with examples of the superposition of two plane waves in nonlinear field theories.
\end{abstract}


\maketitle


\section{\label{sec:Introduction}Introduction\protect\\}

The Skyrme model \cite{Skyrme} is defined by the Lagrangian density
\begin{align}
{\cal L}_S=&-4c_2 \mbox{tr}\left[(g^{\dagger}\partial_{\mu}g)(g^{\dagger}\partial^{\mu}g)\right]+\frac{c_4}{2}\mbox{tr}\left([g^{\dagger}\partial_{\mu}g\,,\,g^{\dagger}\partial_{\nu}g][g^{\dagger}\partial^{\mu}g\,,\,g^{\dagger}\partial^{\nu}g]\right)
\end{align}
where $g(x)$ is an element of $SU(2)$ and $c_2$ and $c_4$ are coupling constants. If we define $A_{\mu}^{\alpha}(x)$ and $H_{\mu\nu}^{\alpha}(x)$ by
\begin{align}
A_{\mu}^{\alpha}=\frac{1}{2i}\mbox{tr}\left(\tau^{\alpha}g^{\dagger}\partial_{\mu}g\right) 
\label{eqn:DofA} \\
 H_{\mu\nu}^{\alpha}=\partial_{\mu}A_{\nu}^{\alpha}-\partial_{\nu}A_{\mu}^{\alpha},
\end{align}
${\cal L}_S$ can be written as
\begin{align}
{\cal L}_S=8c_2A_{\mu}^{\alpha}A^{\alpha, \mu}-c_4H_{\mu\nu}^{\alpha}H^{\alpha, \mu\nu},
\end{align}
where $\tau^{\alpha} \hspace{2mm}(\alpha=1,2,3)$ are the Pauli matrices. The field equation is given by \cite{Skyrme}
\begin{align}
\partial_{\mu}\left(2c_2A^{\alpha, \mu}+c_4\varepsilon^{\alpha\beta\gamma}H^{\beta, \mu\nu}A_{\nu}^{\gamma}\right)=0.
\label{eqn:Motion}
\end{align}
By definition, $A_{\mu}^{\alpha}(x)$ must satisfy the condition
\begin{align}
\partial_{\mu}A_{\nu}^{\alpha}-\partial_{\nu}^{\alpha}A_{\mu}=2\varepsilon^{\alpha\beta\gamma}A_{\mu}^{\beta}A_{\nu}^{\gamma}.
\label{eqn:Condition}
\end{align}
On the other hand, the Faddeev model \cite{FaddeevNiemi,GH,Hietarinta} is defined by the Lagrangian density
\begin{align}
{\cal L}_F=d_2(\partial_{\mu}\mbox{\boldmath$n$})\cdot(\partial^{\mu}\mbox{\boldmath$n$})-4d_4F_{\mu\nu}F^{\mu\nu},
\end{align}
where $\mbox{\boldmath$n$}(x)$ is a three-component vector satisfying
\begin{align}
\mbox{\boldmath$n$}^2=n_an_a=1
\label{eqn:conditionn}
\end{align}
and $F_{\mu\nu}(x)$ is given by
\begin{align}
F_{\mu\nu}=\frac{1}{2}\mbox{\boldmath$n$}\cdot\left(\partial_{\mu}\mbox{\boldmath$n$}\times\partial_{\nu}\mbox{\boldmath $n$}\right).
\label{eqn:DofF}
\end{align}
The field equation for $\mbox{\boldmath$n$}$ turns out to be \cite{Cho}
\begin{align}
\partial_{\mu}\left(d_2\mbox{\boldmath$n$}\times\partial^{\mu}\mbox{\boldmath$n$}-4d_4F^{\mu\nu}\partial_{\nu}\mbox{\boldmath$n$}\right)=0.
\label{eqn:FEQmotion}
\end{align}

Both of these models are expected to describe the effective low energy dynamics of QCD. The soliton solutions of the Skyrme model are identified with baryons \cite{Skyrme}, while those of the Faddeev model are regarded as glueballs \cite{FaddeevNiemi}. The numerical analysis of the Skyrme model generally agrees well with the experimental values \cite{Witten,Jackson}. The further numerical analysis of the Skyrme and the Faddeev models revealed that both models possess knot soliton solutions \cite{Battye,Sutcliffe}.

As for the analytic solutions, however, only a few simple examples are known for these models. Skyrme \cite{Skyrme} pointed out that the field $A_{\mu}^{\alpha}(x)$ of the form
\begin{align}
A_{\mu}^{\alpha}(x)=k_{\mu}f^{\alpha}(k\cdot x)
\label{eqn:SolSkyr}
\end{align}
with $k^2=k_{\mu}k^{\mu}=0$ is a solution of Eq. (\ref{eqn:Motion}) since both  $\partial_{\mu}A^{\alpha, \mu}$ and $H_{\mu\nu}^{\alpha}$ vanish. He also noted that the configuration
\begin{align}
g(x)=\cos{\beta(x)}+ie^{\alpha}\tau^{\alpha}\sin{\beta(x)}
\label{eqn:ConfSkyr}
\end{align}
with $e^{\alpha}\hspace{2mm}(\alpha=1,2,3)$ being real constants satisfying $e^{\alpha}e^{\alpha}=1$ solves the Skyrme model if $\beta(x)$ satisfies $\partial_{\mu}\partial^{\mu}\beta(x)=0$.

The static configuration
\begin{align}
g(x)=\cos\left(\frac{\xi(r)}{2}\right)+\frac{i}{r}x_{\alpha}\tau^{\alpha}\sin\left(\frac{\xi(r)}{2}\right)
\label{eqn:staticSkyr}
\end{align}
with $r=\sqrt{(x_1)^2+(x_2)^2+(x_3)^2}$ is a solution of the Skyrme model if $\xi(r)$ satisfies the differential equation \cite{Cho}
\begin{align}
c_2\left(\frac{d^2\xi}{dr^2}+\frac{2}{r}\frac{d\xi}{dr}-\frac{2\sin\xi}{r^2}\right)+c_4\left[\frac{1-\cos\xi}{r^2}\frac{d^2\xi}{dr^2}+\frac{\sin\xi}{2r^2}\left(\frac{d\xi}{dr}\right)^2-\frac{(1-\cos\xi)\sin^2\xi}{r^4}\right]=0.
\end{align}
It seems difficult to obtain the $c_2, c_4$-dependent analytic solutions of this eqaution for general values of $c_2$ and $c_4$. Recently, Cho \cite{Cho} noted that the constant configuration
\begin{align}
\xi(r)=\pi
\label{eqn:ChoSol}
\end{align}
solves the above equation and constitutes the monopole solution of the Skyrme model. On the other hand, it can be easily checked that the static configuration \cite{Cho}
\begin{align}
n_a(x)=\frac{x_a}{r}
\label{eqn:ChoAnsatz}
\end{align}
is a solution of the Faddeev model since both $\partial_{\mu}(\mbox{\boldmath$n$}\times\partial^{\mu}\mbox{\boldmath$n$})$ and $F^{\mu\nu}$ vanish except at the origin $r=0$. We here remark that all the above solutions do not involve the coupling constants $c_2$ and $c_4$ ($d_2$ and $d_4$) in ${\cal L}_S$ (${\cal L}_F$). The solutions (\ref{eqn:SolSkyr}), (\ref{eqn:ConfSkyr}), and (\ref{eqn:staticSkyr}) with (\ref{eqn:ChoSol}) ((\ref{eqn:ChoAnsatz})) of the Skyrme (Faddeev) model solve the equations $\partial_{\mu}A^{\alpha,\mu}=\partial_{\mu}\left(\varepsilon^{\alpha\beta\gamma}H^{\beta,\mu\nu}A_{\nu}^{\gamma}\right)=0$ ($\partial_{\mu}(\mbox{\boldmath$n$}\times\partial^{\mu}\mbox{\boldmath$n$})=\partial_{\mu}(F^{\mu\nu}\partial_{\nu}\mbox{\boldmath$n$})=0$) which are severer than (\ref{eqn:Motion}) ((\ref{eqn:FEQmotion})). They do not describe the correlation between the $c_2-$ and $c_4-$ ($d_2-$ and $d_4-$) terms in ${\cal L}_S$ (${\cal L}_F$). To see how the two terms in $!
{\cal L}_S$ (${\cal L}_F$) correlate, we must obtain the solutions which depend both on $c_2$ and $c_4$ ($d_2$ and $d_4$). Although, by a suitable choice of the scale of the spacetime and the overall coefficient of ${\cal L}_S$ (${\cal L}_F$), $c_2$ and $c_4$ ($d_2$ and $d_4$) can be set equal to unity, we proceed with keeping them explicit in ${\cal L}_S$ (${\cal L}_F$).

In this paper, we shall obtain a class of exact solutions involving both $c_2$ and $c_4$ ($d_2$ and $d_4$) for the Skyrme (Faddeev) model. With an approriate choice of field variables, the field equations of these two models are written in strictly the same form. Although the final form for $g(x)$ ($\mbox{\boldmath$n$}(x)$) contains the ordering operation and is rather symbolical, the field $A_{\mu}^{\alpha}(x)$ ($\mbox{\boldmath$n$}\times\partial_{\mu}\mbox{\boldmath$n$}$) is given explicitly. The field $A_{\mu}^{\alpha}(x)$ ($\mbox{\boldmath$n$}(x)\times\partial_{\mu}\mbox{\boldmath$n$}(x)$) obtained satisfies the condition for $g(x)$ ($\mbox{\boldmath$n$}(x)$) to exist. We note that many physical quantities, e.g. energy-momentum tensor, can be expressed in terms of $A_{\mu}^{\alpha}(x)$ ($\mbox{\boldmath$n$}(x)\times\partial_{\mu}\mbox{\boldmath$n$}(x)$). They are not the soliton solutions but the wave solutions in the Minkowski space. They supply us with examples how the s!
uperposition of wave solutions is attained in nonlinear field theories. This paper is organized as follows. In Sec. II. we discuss the relation of the field equation of the Faddeev model to that of the Skyrme model. In Sec. III, we solve the Skyrme model and discuss the simplest solution. In Sec. IV, we solve the Faddeev model by the method similar to that of Sec. III. We shall close by a brief summary of our results and an outlook in Sec. V.


 \section{\label{sec:THE RALATION BETWEEN FADDEEV AND SKYRME MODELS}THE RALATION BETWEEN FADDEEV AND SKYRME MODELS\protect\\}

If we define $B_{\mu}^{\alpha}(x)$ by
\begin{align}
B_{\mu}^{\alpha}(x)=\varepsilon^{\alpha\beta\gamma}n^{\beta}(x)\partial_{\mu}n^{\gamma}(x)
\end{align}
and write $\mbox{\boldmath$B$}_{\mu}=(B_{\mu}^1, B_{\mu}^2, B_{\mu}^3)$, we readily obtain the relation
\begin{align}
\mbox{\boldmath$B$}_{\mu}\times\left(\mbox{\boldmath$B$}_{\nu}\times\mbox{\boldmath$B$}_{\rho}\right)=2F_{\nu\rho}\partial_{\mu}\mbox{\boldmath$n$},
\end{align} 
where $F_{\mu\nu}$ is defined by (\ref{eqn:DofF}). Then the field equation of the Faddeev model, (\ref{eqn:FEQmotion}), becomes
\begin{align}
\partial_{\mu}\left[d_2\mbox{\boldmath$B$}^{\mu}+d_4(\mbox{\boldmath$B$}^{\mu}\times\mbox{\boldmath$B$}^{\nu})\times\mbox{\boldmath$B$}_{\nu}\right]=0.
\label{eqn:FaddeevCondition}
\end{align}
By definiton, $\mbox{\boldmath$B$}_{\mu}$ must satisfy the condition
\begin{align}
\partial_{\mu}\mbox{\boldmath$B$}_{\nu}-\partial_{\nu}\mbox{\boldmath$B$}_{\mu}=2\mbox{\boldmath$B$}_{\mu}\times\mbox{\boldmath$B$}_{\nu}.
\label{eqn:FaddeevMotion}
\end{align}
We find that Eqs. (\ref{eqn:FaddeevCondition}) and (\ref{eqn:FaddeevMotion}) are strictly of the same form as Eqs. (\ref{eqn:Motion}) and (\ref{eqn:Condition}). Since the degree of freedom in $\mbox{\boldmath$n$}(x)$ is two while that in $g(x)$ is three, we conclude that the solutions of the Faddeev model yield a restricted class of solutions of the Skyrme model.


 \section{\label{sec:Solution of the Skyrme model}Solution of the Skyrme model\protect\\}
\subsection{\label{sec:Solution}Solution\protect\\}
In this section, we seek the solution of Eq. (\ref{eqn:Motion}) of the following form:
\begin{align}
&g(x)=h(\xi,\eta), \\
&\xi=k\cdot x,\hspace{2mm} \eta=l\cdot x,\nonumber
\end{align}
where $k$ and $l$ are constant four-vectors satisfying 
\begin{align}
k^2=l^2=0,\hspace{2mm} k\cdot l\neq 0.
\label{eqn:kl0}
\end{align}
It can be regarded as a generalization of (\ref{eqn:SolSkyr}). The field $A_{\mu}^{\alpha}(x)$ defined by Eq. (\ref{eqn:DofA}) is written as
\begin{align}
&A_{\mu}^{\alpha}(x)=a^{\alpha}(\xi,\eta)k_{\mu}+b^{\alpha}(\xi,\eta)l_{\mu}, \nonumber\\
&a^{\alpha}=\frac{1}{2i}\mbox{tr}\left[\tau^{\alpha}h^{\dagger}(\xi, \eta)\frac{\partial h(\xi, \eta)}{\partial\xi}\right], \\
&b^{\alpha}=\frac{1}{2i}\mbox{tr}\left[\tau^{\alpha}h^{\dagger}(\xi, \eta)\frac{\partial h(\xi, \eta)}{\partial\eta}\right].\nonumber
\end{align}
In terms of the vectors $\mbox{\boldmath$a$}=(a^1,a^2,a^3)$ and $\mbox{\boldmath$b$}=(b^1,b^2,b^3)$, the field equation (\ref{eqn:Motion}) and the conditon (\ref{eqn:Condition}) are expressed as
\begin{align}
&\frac{\partial}{\partial\eta}\left[\sigma\mbox{\boldmath$a$}+\mbox{\boldmath$a$}\times(\mbox{\boldmath$b$}\times\mbox{\boldmath$a$})\right]+\frac{\partial}{\partial\xi}\left[\sigma\mbox{\boldmath$b$}+\mbox{\boldmath$b$}\times(\mbox{\boldmath$a$}\times\mbox{\boldmath$b$})\right]=0,
\label{eqn:EQab} \\
&\frac{\partial\mbox{\boldmath$a$}}{\partial\eta}-\frac{\partial\mbox{\boldmath$b$}}{\partial\xi}=2(\mbox{\boldmath$b$}\times\mbox{\boldmath$a$}),
\label{eqn:Conditionab}
\end{align}
where $\sigma$ is defined by
\begin{align}
\sigma=\frac{c_2}{c_4(k\cdot l)}.
\label{eqn:Dofsigma}
\end{align}
If we assume that $\mbox{\boldmath$b$}$ is a constant vector, Eqs. (\ref{eqn:EQab}) and (\ref{eqn:Conditionab}) are simplified to
\begin{align}
&\frac{\partial\mbox{\boldmath$a$}}{\partial\eta}=2(\mbox{\boldmath$b$}\times\mbox{\boldmath$a$}), 
\label{eqn:atimesb}\\
&\mbox{\boldmath$b$}\times\left\{\mbox{\boldmath$b$}\times\frac{\partial\mbox{\boldmath$a$}}{\partial\xi}+2\left[(\mbox{\boldmath$a$}\cdot\mbox{\boldmath$b$})-\sigma\right]\mbox{\boldmath$a$}\right\}=0.
\label{eqn:FEQfinalS}
\end{align}
The latter equation is equivalent to
\begin{align}
\mbox{\boldmath$b$}\times\left\{\frac{\partial\mbox{\boldmath$a$}}{\partial\xi}+\frac{2\left[(\mbox{\boldmath$a$}\cdot\mbox{\boldmath$b$})-\sigma\right]}{b^2}(\mbox{\boldmath$a$}\times\mbox{\boldmath$b$})\right\}=0,
\label{eqn:EQAB}
\end{align}
where $b$ is equal to $\sqrt{\mbox{\boldmath$b$}^2}$. From (\ref{eqn:EQAB}), we have
\begin{align}
\frac{\partial\mbox{\boldmath$a$}}{\partial\xi}=\frac{2[(\mbox{\boldmath$a$}\cdot\mbox{\boldmath$b$})-\sigma]}{b^2}(\mbox{\boldmath$a$}\times\mbox{\boldmath$b$})+{\cal K}(\xi,\eta)\mbox{\boldmath$b$},
\label{eqn:Diffaxi}
\end{align}
where ${\cal K}(\xi, \eta)$ is an arbitrary function of $\xi$ and $\eta$. Then from (\ref{eqn:atimesb}) and (\ref{eqn:Diffaxi}), we obtain
\begin{align}
\frac{\partial}{\partial\xi}\left(\frac{\partial\mbox{\boldmath$a$}}{\partial\eta}\right)=\frac{4[(\mbox{\boldmath$a$}\cdot\mbox{\boldmath$b$})-\sigma]}{b^2}\left[\mbox{\boldmath$b$}\times(\mbox{\boldmath$b$}\times\mbox{\boldmath$a$})\right]
\label{eqn:dxidzeta}
\end{align} 
and
\begin{align}
\frac{\partial}{\partial\eta}\left(\frac{\partial\mbox{\boldmath$a$}}{\partial\xi}\right)=\frac{4[(\mbox{\boldmath$a$}\cdot\mbox{\boldmath$b$})-\sigma]}{b^2}\left[\mbox{\boldmath$b$}\times(\mbox{\boldmath$b$}\times\mbox{\boldmath$a$})\right]+\frac{\partial{\cal K}(\xi, \eta)}{\partial\eta}\mbox{\boldmath$b$}.
\label{eqn:detadxi}
\end{align}
In deriving (\ref{eqn:detadxi}), we have made use of the result of (\ref{eqn:atimesb}) that $\mbox{\boldmath$a$}\cdot\mbox{\boldmath$b$}$ is independent of $\eta$. Comparing (\ref{eqn:dxidzeta}) with (\ref{eqn:detadxi}), we find that ${\cal K}(\xi, \eta)$ is independent of $\eta$ if $\mbox{\boldmath$a$}$ is a regular function of $\xi$ and $\eta$. ${\cal K}(\xi, \eta)$ in (\ref{eqn:Diffaxi}) should be replaced by ${\cal K}(\xi)$:
\begin{align}
\frac{\partial\mbox{\boldmath$a$}}{\partial\xi}=\frac{2[(\mbox{\boldmath$a$}\cdot\mbox{\boldmath$b$})-\sigma]}{b^2}\left(\mbox{\boldmath$a$}\times\mbox{\boldmath$b$}\right)+{\cal K}(\xi)\mbox{\boldmath$b$},
\label{eqn:Kxi}
\end{align}
from which we have
\begin{align}
\frac{\partial}{\partial\xi}\left(\mbox{\boldmath$a$}\cdot\mbox{\boldmath$b$}\right)={\cal K}(\xi)b^2.
\end{align}
If we define $\nu(\xi)$ by
\begin{align}
\nu(\xi)=\frac{\left(\mbox{\boldmath$a$}\cdot\mbox{\boldmath$b$}\right)-\sigma}{b^2},
\label{eqn:Dofnu}
\end{align}
${\cal K}(\xi)$ is given by
\begin{align}
{\cal K}(\xi)=\frac{d\nu(\xi)}{d\xi}.
\label{eqn:DofK}
\end{align}
Eqs. (\ref{eqn:Kxi}), (\ref{eqn:Dofnu}) and (\ref{eqn:DofK}) now yield
\begin{align}
\frac{\partial\mbox{\boldmath$a$}}{\partial\xi}=2\nu(\xi)\left(\mbox{\boldmath$a$}\times\mbox{\boldmath$b$}\right)+\frac{d\nu(\xi)}{d\xi}\mbox{\boldmath$b$}.
\end{align}

If we define $\mbox{\boldmath$p$}(\xi, \eta)$ and $\omega(\xi)$ by
\begin{align}
&\mbox{\boldmath$p$}(\xi,\eta)=\mbox{\boldmath$a$}(\xi,\eta)-\nu(\xi)\mbox{\boldmath$b$}, 
\label{eqn:Dofp} \\
&\frac{d\omega(\xi)}{d\xi}=\nu(\xi),
\label{eqn:Dofomega}
\end{align}
we are led to the simple equations
\begin{align}
\frac{\partial\mbox{\boldmath$p$}}{\partial\eta}=\frac{\partial\mbox{\boldmath$p$}}{\partial\omega}=2\left(\mbox{\boldmath$b$}\times\mbox{\boldmath$p$}\right).
\end{align}
The general solution of the above equation is given by
\begin{align}
\mbox{\boldmath$p$}(\xi,\eta)=\lambda\mbox{\boldmath$b$}+\mbox{\boldmath$c$}\cos\left[2b(\eta+\omega(\xi))\right]+\mbox{\boldmath$d$}\sin\left[2b(\eta+\omega(\omega))\right],
\label{eqn:generalp}
\end{align}
where $\lambda$ is a constant and $\mbox{\boldmath$c$}$ and $\mbox{\boldmath$d$}$ are constant vectors satisfying
\begin{align}
\mbox{\boldmath$b$}\times\mbox{\boldmath$c$}=b\mbox{\boldmath$d$}, \hspace{2mm} \mbox{\boldmath$d$}\times\mbox{\boldmath$b$}=b\mbox{\boldmath$c$}.
\end{align}
Comparing the result (\ref{eqn:generalp}) with Eqs. (\ref{eqn:Dofnu}), (\ref{eqn:Dofp}) and (\ref{eqn:Dofomega}), we finally obtain $\lambda=\sigma/b^2$ and
\begin{align}
\mbox{\boldmath$a$}(\xi,\eta)&=\left[\frac{\sigma}{b^2}+\frac{d\omega(\xi)}{d\xi}\right]\mbox{\boldmath$b$}+\mbox{\boldmath$c$}\cos\left[2b(\eta+\omega(\xi))\right]+\mbox{\boldmath$d$}\sin\left[2b(\eta+\omega(\xi))\right]
\label{eqn:repofa}
\end{align}
with an arbitrary function $\omega(\xi)$. We observe that the nonlinearity in $\mbox{\boldmath$a$}$ of the l.h.s. of Eq. (\ref{eqn:EQAB}) has been absorbed in the definition of the variable $\omega(\xi)$: $\mbox{\boldmath$a$}\cdot\mbox{\boldmath$b$}=b^2\frac{d\omega(\xi)}{d\xi}+\sigma$. We note that this condition, Eq.(\ref{eqn:Dofnu}), is automatically satisfied for $\mbox{\boldmath$a$}$ given by (\ref{eqn:repofa}). It is easy to check that $\mbox{\boldmath$a$}(\xi, \eta)$ given by Eq. (\ref{eqn:repofa}) indeed solves Eq. (\ref{eqn:EQAB}). Thus we have
\begin{align}
&A_{\mu}^{\alpha}(x)=\left\{k_{\mu}\left[\frac{\sigma}{b^2}+\frac{d\omega(\xi)}{d\xi}\right]+l_{\mu}\right\}b^{\alpha}+k_{\mu}\left\{c^{\alpha}\cos\left[2b(\eta+\omega(\xi))\right]+d^{\alpha}\sin\left[2b(\eta+\omega(\xi))\right]\right\}, 
\label{eqn:LongA} \\
&H_{\mu\nu}^{\alpha}(x)=\left(l_{\mu}k_{\nu}-l_{\nu}k_{\mu}\right)\left\{-c^{\alpha}\sin\left[2b(\eta+\omega(\xi))\right]+d^{\alpha}\cos\left[2b(\eta+\omega(\xi))\right]\right\}, \\
&\frac{d\omega(\xi)}{d\xi}=-\frac{\sigma}{b^2}+\frac{1}{l\cdot k}b^{\alpha}l^{\mu}A_{\mu}^{\alpha}(x).
\end{align}

We see that the coupling constants $c_2$ and $c_4$ appear in $A_{\mu}^{\alpha}(x)$ through $\sigma$ and possibly through $\omega(\xi)$ and that $\omega(\xi)$ is determined by a certain condition imposed on $b^{\alpha}l^{\mu}A_{\mu}^{\alpha}(x)$.

We note that, for the above solution, we have the relation
\begin{align}
&\partial_{\mu}A^{\alpha, \mu}=\frac{2b}{(k\cdot l)}k^{\mu}l^{\nu}H_{\mu\nu}^{\alpha} \nonumber \\
&=2b(k\cdot l)\left\{-c^{\alpha}\sin\left[2b(\eta+\omega(\xi))\right]+d^{\alpha}\cos\left[2b(\eta+\omega(\xi))\right]\right\}.
\end{align}

We next obtain $g(x)=h(\xi,\eta)$ by assuming
\begin{align}
h(\xi,\eta)=u(\xi)v(\eta),
\label{eqn:Dofh}
\end{align}
where $u(\xi)$ and $v(\eta)$ belong to $SU(2)$. From Eqs. (\ref{eqn:DofA}) and (\ref{eqn:Dofh}), we have
\begin{align}
&a^{\alpha}=\frac{1}{2i}\mbox{tr}\left(\tau^{\alpha}v^{\dagger}u^{\dagger}\frac{du}{d\xi}v\right),\nonumber \\
&b^{\alpha}=\frac{1}{2i}\mbox{tr}\left(\tau^{\alpha}v^{\dagger}\frac{dv}{d\eta}\right).
\end{align} 
$v(\eta)$ is now given by
\begin{align}
v(\eta)&=e^{i\eta\mbox{\boldmath$b$}\cdot\mbox{\boldmath$\tau$}}, \nonumber\\
&=\cos(\eta b)+i(\hat{\mbox{\boldmath$b$}}\cdot\mbox{\boldmath$\tau$})\sin(\eta b)
\end{align}
with $\hat{\mbox{\boldmath$b$}}=\mbox{\boldmath$b$}/b$. Then $v \tau^{\alpha}v^{\dagger}$ is calculated to be
\begin{align}
&v\tau^{\alpha}v^{\dagger}=m^{\alpha\beta}\tau^{\beta}, \nonumber \\
&m^{\alpha\beta}=\left(\delta^{\alpha\beta}-\hat{b}^{\alpha}\hat{b}^{\beta}\right)\cos(2b\eta)-\varepsilon^{\alpha\beta\gamma}\hat{b}^{\gamma}\sin(2b\eta)+\hat{b}^{\alpha}\hat{b}^{\beta}.
\end{align}
Comparing the expressions
\begin{align}
a^{\alpha}=\frac{m^{\alpha\beta}}{2i}\mbox{tr}\left(\tau^{\beta}u^{\dagger}\frac{du}{d\xi}\right)
\end{align}
with the result (\ref{eqn:repofa}), we obtain
\begin{align}
&\frac{1}{2i}\left(\delta^{\alpha\beta}-\hat{b}^{\alpha}\hat{b}^{\beta}\right)\mbox{tr}\left(\tau^{\beta}u^{\dagger}\frac{du}{d\xi}\right)=c^{\alpha}\cos(2b\omega)+d^{\alpha}\sin(2b\omega), \nonumber\\
&-\frac{1}{2i}\varepsilon^{\alpha\beta\gamma}\hat{b}^{\gamma}\mbox{tr}\left(\tau^{\beta}u^{\dagger}\frac{du}{d\xi}\right)=-c^{\alpha}\sin(2b\omega)+d^{\alpha}\cos(2b\omega), \nonumber\\
&\frac{1}{2i}\hat{b}^{\alpha}\hat{b}^{\beta}\mbox{tr}\left(\tau^{\beta}u^{\dagger}\frac{du}{d\xi}\right)=b\left[\frac{\sigma}{b^2}+\frac{d\omega(\xi)}{d\xi}\right]\hat{b}^{\alpha}.
\end{align}
The solution of these equations is given by
\begin{align}
\frac{1}{2i}\mbox{tr}\left(\tau^{\beta}u^{\dagger}\frac{du}{d\xi}\right)=q^{\beta}(\xi),
\end{align}
where $q^{\alpha}(\xi)$ is defined by
\begin{align}
q^{\alpha}(\xi)=b^{\alpha}\left[\frac{\sigma}{b^2}+\frac{d\omega(\xi)}{d\xi}\right]+c^{\alpha}\cos[2b\omega(\xi)]+d^{\alpha}\sin[2b\omega(\xi)].
\end{align}
Now $u(\xi)$ is obtained as
\begin{align}
u(\xi)=u(0)\overline{P_{\xi'}}\exp\left[i\int_0^{\xi}d\xi'\tau^{\alpha}q^{\alpha}(\xi')\right],
\label{eqn:Solu}
\end{align}
where $\overline{P_{\xi'}}$ denotes the anti-$\xi'$-ordering.

\subsection{\label{sec:Discussion}Discussion\protect\\}

For general $\omega(\xi)$, the r.h.s. of (\ref{eqn:Solu}) cannot be integrated explicitly. We can, however, see that the above result implies some nontrivial nonlinear effect caused by the interplay of the $c_2-$ and $c_4-$ terms in ${\cal L}_S$. For example, in the simplest case that $\omega(\xi)$ is a constant, $q^{\alpha}(\alpha=1,2,3)$ are constants also and we have
\begin{align}
&g(x)=e^{i\xi\mbox{\boldmath$q$}\cdot\mbox{\boldmath$\tau$}}e^{i\eta\mbox{\boldmath$b$}\cdot\mbox{\boldmath$\tau$}}, \\
&\mbox{\boldmath$q$}=\mbox{\boldmath$b$}\frac{\sigma}{b^2}+\mbox{\boldmath$c$}\cos(2b\omega)+\mbox{\boldmath$d$}\sin(2b\omega).
\end{align}
The two vectors $\mbox{\boldmath$q$}$ and $\mbox{\boldmath$b$}$ defining $g(x)$ satisfy
\begin{align}
\mbox{\boldmath$q$}\cdot\mbox{\boldmath$b$}=\sigma.
\end{align}
It is straightfoward to see that $g(x)$ a superposition of the waves described by $\sin[j_{\pm}(x)]$ and $\cos[j_{\pm}(x)]$, where $j_{+}(x)$ and $j_{-}(x)$ are defined by
\begin{align}
j_{\pm}(x)=\left[k\sqrt{\left(\frac{c_2}{c_4}\right)^2\frac{1}{(k\cdot l)^2}\frac{1}{b^2}+c^2}\pm bl\right]\cdot x\equiv k_{\pm}\cdot x,
\end{align}
where we have made use of (\ref{eqn:Dofsigma}) and $\mbox{\boldmath$c$}^2=\mbox{\boldmath$b$}^2\equiv c^2$. We observe that the above type of solutions would never appear in the model defined by ${\cal L}_S$ with $c_4=0$. It is interesting that $\left[(k_{\pm})^2\right]^2$ is given by
\begin{align}
\left[(k_{\pm})^2\right]^2=\left(\frac{c_2}{c_4}\right)^2+b^2c^2.
\end{align}

We next consider the energy-momentum tensor associated with the solution (\ref{eqn:Solu}). Since the energy-momentum tensor
\begin{align}
T_{\mu\nu}=\frac{\partial{\cal L}_S}{\partial A^{\alpha,\mu}}A_{\nu}^{\alpha}-\eta_{\mu\nu}{\cal L}_S
\label{eqn:EMtensor}
\end{align}
can be expressed in terms of $A_{\mu}^{\alpha}$ \cite{Skyrme}, we are not worried about the ordering operation $\overline{P}_{\xi'}$ in (\ref{eqn:Solu}). From the results (\ref{eqn:LongA}) and (\ref{eqn:EMtensor}), we have
\begin{align}
T_{\mu\nu}=&16c_2\left[(z^2b^2+c^2)k_{\mu}k_{\nu}+zb^2(k_{\mu}l_{\nu}+k_{\nu}l_{\mu})+b^2l_{\mu}l_{\nu}-(k\cdot l)zb^2\eta_{\mu\nu}\right] \nonumber \\
&+2c_4(k\cdot l)c^2\left[8b^2(k_{\mu}l_{\nu}+k_{\nu}l_{\mu})-(k\cdot l)\eta_{\mu\nu}\right], \\
z=&\frac{\sigma}{b^2}+\frac{d\omega(\xi)}{d\xi}.
\end{align}
We see that $T_{\mu\nu}$ depend only on $\xi$. From the above $T_{\mu\nu}$, we obtain
\begin{align}
&T_{\mu}^{\mu}=-32c_2zb^2(k\cdot l)+8c_4(k\cdot l)^2c^2(4b^2-1), \\
&k^{\mu}T_{\mu\nu}=16c_2b^2(k\cdot l)l_{\nu}+2c_4(k\cdot l)^2c^2(8b^2-1)k_{\nu}, \\
&l^{\mu}T_{\mu\nu}=16c_2(z^2b^2+c^2)(k\cdot l)k_{\nu}+2c_4(k\cdot l)^2c^2(8b^2-1)l_{\nu}.
\end{align}
We find that $k^{\mu}T_{\mu\nu}$ is constant, while $l^{\mu}T_{\mu\nu}$ depends on $\xi$.

In contrast to the above results, in the case of the solution (\ref{eqn:SolSkyr}), we have $T_{\mu\nu}=16c_2k_{\mu}k_{\nu}f^{\alpha}(k\cdot x)f^{\alpha}(k\cdot x)$ and hence $T_{\mu}^{\mu}=k^{\mu}T_{\mu\nu}=0$.


 \section{\label{sec:SOLUTIONS OF THE FADDEEV MODEL}SOLUTIONS OF THE FADDEEV MODEL \protect\\}

We next obtain the solution of the Faddeev model. We assume that $n_a(x)\hspace{2mm} (a=1,2,3)$ are functions of the variables $\xi=k\cdot x$ and $\eta=l\cdot x$ introduced in the previous section. We still adopt the assumption (\ref{eqn:kl0}) for $k_{\mu}$ and $l_{\mu}$. \\
Then we have
\begin{align}
F_{\mu\nu}=\frac{1}{2}\left(k_{\mu}l_{\nu}-k_{\nu}l_{\mu}\right)\left[\mbox{\boldmath$n$}\cdot\left(\frac{\partial\mbox{\boldmath$n$}}{\partial\xi}\times\frac{\partial\mbox{\boldmath$n$}}{\partial\xi}\right)\right].
\label{eqn:Fmunu}
\end{align}
The field equation (\ref{eqn:FEQmotion}) becomes
\begin{align}
&\left(k_{\mu}\frac{\partial}{\partial\xi}+l_{\mu}\frac{\partial}{\partial\eta}\right)\left\{d_2\left[k^{\mu}\left(\mbox{\boldmath$n$}\times\frac{\partial\mbox{\boldmath$n$}}{\partial\xi}\right)+l^{\mu}\left(\mbox{\boldmath$n$}\times\frac{\partial\mbox{\boldmath$n$}}{\partial\eta}\right)\right]\right. \nonumber \\
&\left.-2d_4\left(k^{\mu}l^{\nu}-k^{\nu}l^{\mu}\right)\left(k_{\nu}\frac{\partial\mbox{\boldmath$n$}}{\partial\xi}+l_{\nu}\frac{\partial\mbox{\boldmath$n$}}{\partial\eta}\right)\left[\mbox{\boldmath$n$}\cdot\left(\frac{\partial\mbox{\boldmath$n$}}{\partial\xi}\times\frac{\partial\mbox{\boldmath$n$}}{\partial\eta}\right)\right]\right\}=0.
\label{eqn:FEQlong}
\end{align}
Guided by the discussion in Sec. II, we define $\mbox{\boldmath$\alpha$}$ and $\mbox{\boldmath$\beta$}$ by
\begin{align}
&\mbox{\boldmath$\alpha$}=\mbox{\boldmath$n$}\times\frac{\partial\mbox{\boldmath$n$}}{\partial\xi},
\label{eqn:Dofalpha} \\
&\mbox{\boldmath$\beta$}=\mbox{\boldmath$n$}\times\frac{\partial\mbox{\boldmath$n$}}{\partial\eta}.
\label{eqn:Dofbeta}
\end{align}
If we take the condition (\ref{eqn:conditionn}) into account, we have
\begin{align}
\mbox{\boldmath$\alpha$}\times\mbox{\boldmath$\beta$}=\frac{\partial\mbox{\boldmath$n$}}{\partial\xi}\times\frac{\partial\mbox{\boldmath$n$}}{\partial\eta}.
\end{align}
Then the field equation (\ref{eqn:FEQlong}) is written as
\begin{align}
\frac{\partial}{\partial\eta}\left[\rho \mbox{\boldmath$\alpha$}+\mbox{\boldmath$\alpha$}\times(\mbox{\boldmath$\beta$}\times\mbox{\boldmath$\alpha$})\right]+\frac{\partial}{\partial\xi}\left[\rho \mbox{\boldmath$\beta$}+\mbox{\boldmath$\beta$}\times(\mbox{\boldmath$\alpha$}\times\mbox{\boldmath$\beta$})\right]=0,
\label{eqn:FEQshort}
\end{align}
where $\rho$ is defined by
\begin{align}
\rho=\frac{d_2}{2d_4(l\cdot k)}.
\end{align}
On the other hand, the definitions (\ref{eqn:Dofalpha}) and (\ref{eqn:Dofbeta}) yield
\begin{align}
\frac{\partial\mbox{\boldmath$\alpha$}}{\partial\eta}-\frac{\partial\mbox{\boldmath$\beta$}}{\partial\xi}=2(\mbox{\boldmath$\beta$}\times\mbox{\boldmath$\alpha$}).
\label{eqn:conditionalpha}
\end{align}
We find that the above equations for $\mbox{\boldmath$\alpha$}$ and $\mbox{\boldmath$\beta$}$ are strictry of the same form as those for $\mbox{\boldmath$a$}$ and $\mbox{\boldmath$b$}$ in the previous section. If the solutions $\mbox{\boldmath$a$}$ and $\mbox{\boldmath$b$}$ for the Skyrme model are given by $\mbox{\boldmath$a$}=\mbox{\boldmath$F$}(\xi,\eta;\sigma)$ and $\mbox{\boldmath$b$}=\mbox{\boldmath$G$}(\xi,\eta;\sigma)$, then $\mbox{\boldmath$\alpha$}=\mbox{\boldmath$F$}(\xi,\eta;\rho)$ and $\mbox{\boldmath$\beta$}=\mbox{\boldmath$G$}(\xi,\eta;\rho)$ solve Eqs. (\ref{eqn:FEQshort}) and (\ref{eqn:conditionalpha}). If we assume that $\mbox{\boldmath$\beta$}$ is a constant vector, we obtain the analogue of Eqs. (\ref{eqn:atimesb}), (\ref{eqn:FEQfinalS}) and (\ref{eqn:Dofnu}):
\begin{align}
&\frac{\partial\mbox{\boldmath$\alpha$}}{\partial\eta}=2(\mbox{\boldmath$\beta$}\times\mbox{\boldmath$\alpha$}), \nonumber\\
&\frac{\partial\mbox{\boldmath$\alpha$}}{\partial\xi}=2\mu(\xi)\left(\mbox{\boldmath$\beta$}\times\mbox{\boldmath$\alpha$}\right)+\frac{d\mu(\xi)}{d\xi}\mbox{\boldmath$\beta$}, \nonumber\\
&\mu(\xi)=\frac{(\mbox{\boldmath$\alpha$}\cdot\mbox{\boldmath$\beta$})-\rho}{\beta^2}
\end{align}
with $\beta=\sqrt{\mbox{\boldmath$\beta$}^2}$.

We now know that there exists $\tilde{g}(x)=\tilde{h}(\xi,\eta)=\tilde{v}(\xi)\tilde{u}(\eta),\tilde{g},\tilde{h},\tilde{u},\tilde{v}\in SU(2)$, such that
\begin{align}
&A\equiv i\alpha^{\alpha}\tau^{\alpha}=\tilde{g}^{\dagger}\frac{\partial\tilde{g}}{\partial\xi}=\tilde{v}^{\dagger}\tilde{u}^{\dagger}\frac{d\tilde{u}}{d\xi}\tilde{v}, \\
&B\equiv i\beta^{\alpha}\tau^{\alpha}=\tilde{g}^{\dagger}\frac{\partial\tilde{g}}{\partial\eta}=\tilde{v}^{\dagger}\frac{d\tilde{v}}{d\eta}.
\end{align}
They are obtained by replacing $\sigma$ in $g,h,u$ and $v$ of Sec. III by $\rho$. The fields $n_a$ are obtained from Eqs. (\ref{eqn:Dofalpha}) and (\ref{eqn:Dofbeta}) or equivalently from
\begin{align}
&\frac{\partial\mbox{\boldmath$n$}}{\partial\xi}=\mbox{\boldmath$\alpha$}\times\mbox{\boldmath$n$}, 
\label{eqn:Repalpha} \\
&\frac{\partial\mbox{\boldmath$n$}}{\partial\eta}=\mbox{\boldmath$\beta$}\times\mbox{\boldmath$n$}.
\label{eqn:Repbeta}
\end{align} 
If we define $N$ by
\begin{align}
N=\tau^an_a,
\end{align}
Eqs. (\ref{eqn:Repalpha}) and (\ref{eqn:Repbeta}) become
\begin{align}
\frac{\partial N}{\partial\xi}=\frac{1}{2}\left[N, A\right],
\label{eqn:dndxi} \\
\frac{\partial N}{\partial\eta}=\frac{1}{2}\left[N, B\right].
\label{eqn:dndeta}
\end{align}
$\mbox{\boldmath$n$}$ is determined by these conditions in the following way. We set $N$ as
\begin{align}
N=j^{\dagger}qj,
\end{align}
where $j$ belongs to $SU(2)$ and $q$ is a constant matrix belonging to $su(2)$. The condition (\ref{eqn:conditionn}) yields
\[
q^2=\left(\begin{array}{cc}
1&0 \\
0&1 \\
\end{array}\right),
\]
while Eqs. (\ref{eqn:dndxi}) and (\ref{eqn:dndeta}) give
\begin{align}
\left[N, j^{\dagger}\frac{\partial j}{\partial\xi}\right]=\frac{1}{2}\left[N, \tilde{g}^{\dagger}\frac{\partial\tilde{g}}{\partial\xi}\right], \\
\left[N, j^{\dagger}\frac{\partial j}{\partial\eta}\right]=\frac{1}{2}\left[N, \tilde{g}^{\dagger}\frac{\partial\tilde{g}}{\partial\eta}\right].
\end{align}
If we define the analogue of $q^{\alpha}(\xi)$ of Sec. III by $\tilde{q}^{\alpha}(\xi)$, that is,
\begin{align}
&\tilde{q}^{\alpha}(\xi)=\beta^{\alpha}\left[\frac{\rho}{\beta^2}+\frac{d\omega(\xi)}{d\xi}\right]+\gamma^{\alpha}\cos[2\beta\omega(\xi)]+\delta^{\alpha}\sin[2\beta\omega(\xi)], \\
&\mbox{\boldmath$\beta$}\times\mbox{\boldmath$\gamma$}=\beta\mbox{\boldmath$\delta$},\hspace{5mm}\mbox{\boldmath$\delta$}\times\mbox{\boldmath$\beta$}=\beta\mbox{\boldmath$\gamma$},
\end{align}
we have
\begin{align}
&j^{\dagger}\frac{\partial j}{\partial\xi}=\tilde{v}^{\dagger}\left(\frac{i}{2}\tau^{\alpha}\tilde{q}^{\alpha}(\xi)\right)\tilde{v} \\
&j^{\dagger}\frac{\partial j}{\partial\eta}=\tilde{v}^{\dagger}\left(\frac{i}{2}\tau^{\alpha}\beta^{\alpha}\right)\tilde{v}=\frac{i}{2}{\tau}^{\alpha}\beta^{\alpha}.
\end{align}
We then obtain
\begin{align}
&j(\xi,\eta)=j(0,0)\overline{P_t}\exp\left[i\int^1_0dt\Gamma(t)\right], \\
&\Gamma(t)=\frac{1}{2}\tilde{v}^{\dagger}\left[\zeta(t)\right]\tau^{\alpha}\left\{\tilde{q}^{\alpha}\left[\xi(t)\right]\frac{d\xi(t)}{dt}+\beta^{\alpha}\frac{d\zeta(t)}{dt}\right\}\tilde{v}\left[\zeta(t)\right]
\end{align}
with
\begin{align}
\xi(0)=\eta(0)=0,\hspace{5mm}\xi(1)=\xi,\hspace{5mm}\eta(1)=\eta.
\end{align}
As in the case of the Skyrme model, we can calculate some quantities without being worried by the ordering operator $\overline{P}_t$. For example, from Eqs. (\ref{eqn:Fmunu}), (\ref{eqn:Dofalpha}), (\ref{eqn:Dofbeta}), and the discussion below (\ref{eqn:conditionalpha}), we have
\begin{align}
F_{\mu\nu}F^{\mu\nu}&=-\frac{1}{2}(k\cdot l)^2\left(\mbox{\boldmath$\alpha$}\times\mbox{\boldmath$\beta$}\right)^2 \nonumber\\
&=\left.-\frac{1}{2}(k\cdot l)^2\left(\mbox{\boldmath$a$}\times\mbox{\boldmath$b$}\right)^2\right|_{\sigma\rightarrow\rho}\nonumber \\
&=-\frac{1}{2}(k\cdot l)^2b^2c^2,
\end{align}
which is constant. The term $\partial_{\mu}\mbox{\boldmath$n$}\cdot\partial^{\mu}\mbox{\boldmath$n$}$ in ${\cal L}_F$ is equal to $2(k\cdot l)(\mbox{\boldmath$\alpha$}\cdot\mbox{\boldmath$\beta$})=2(k\cdot l)\left(\rho+\beta^2\frac{d\omega(\xi)}{d\xi}\right)$.


\section{\label{sec:Summary And Outlook} SUMMARY AND OUTLOOK\protect\\}

We have discussed the intimate relation between the Faddeev and the Skyrme models. We have seen that the field equations and the supplementary conditions for these models take exactly the same form. Assuming that $g(x)$ in ${\cal L}_S$ and $\mbox{\boldmath$n$}(x)$ in ${\cal L}_F$ are functions of the variables $\xi=k\cdot x$ and $\eta=l\cdot x$ with $k^2=l^2=0$, we have obtained a class of exact solutions of the Faddeev and the Skyrme models. In contrast with the previously obtained solutions of these models, our solutions depend on two coupling constants which define the models. Our solutions contain an arbitrary function $\omega(\xi)$.

The baryon number current of the Skyrme model is defined by \cite{Skyrme}
\begin{align}
N^{\lambda}=\frac{1}{12\pi^2}\varepsilon^{\lambda\mu\nu\rho}\varepsilon^{\alpha\beta\gamma}A_{\mu}^{\alpha}A_{\nu}^{\beta}A_{\rho}^{\gamma}
\end{align} 
and vanishes under our assumption. Our results, however, suggests that there would be a wave solution of the Skyrme model which is a function of $\xi , \eta,$ and $\zeta=m\cdot x$ with $m^2=0$. For such a solution, although the baryon number $N=\int d^3x N_0(x)$ might be ill-defined, the density $N_0(x)$ itself would take a well-defined nonvanishing value.


\begin{acknowledgments}
The authors are grateful to Shinji Hamamoto, Takeshi Kurimoto, Hiroshi Kakuhata, Hitoshi Yamakoshi, Hideaki Hayakawa and Chang-Guang Shi for discussions.
\end{acknowledgments}

\end{document}